\documentclass[
showpacs,
floatfix,
aps,
prl,
twocolumn,
superscriptaddress,
amssymb,
]{revtex4-1}

\usepackage{amssymb}
\usepackage{latexsym}
\usepackage[dvips]{graphicx}
\usepackage{amsmath}
\usepackage{float}

\usepackage{graphicx}
\usepackage{dcolumn}
\usepackage{amsfonts}
\usepackage{bm}

\newcommand{\be}{\begin{equation}}
\newcommand{\ee}{\end{equation}}
\newcommand{\bea}{\begin{eqnarray}}
\newcommand{\eea}{\end{eqnarray}}

\def\bip{B_\parallel}

\def\rxx{R_{xx}}
\def\ryy{R_{yy}}

\def\easy{\left < 110 \right >}

\def\x{\hat{x}}
\def\y{\hat{y}}
\def\easy{\left < 110 \right >}

\def\A{A_R}

\newcommand{\rfig}[1]{Fig.\,\ref{#1}}

\newcommand{\rref}[1]{Ref.\,\onlinecite{#1}}

\begin{document}
\title{Reorientation of quantum Hall stripes within a partially filled Landau level}
\author{Q.~Shi}
\affiliation{School of Physics and Astronomy, University of Minnesota, Minneapolis, Minnesota 55455, USA}
\author{M.~A.~Zudov}
\email[Corresponding author: ]{zudov@physics.umn.edu}
\affiliation{School of Physics and Astronomy, University of Minnesota, Minneapolis, Minnesota 55455, USA}
\author{J.\,D. Watson}
\affiliation{Department of Physics and Astronomy, Purdue University, West Lafayette, Indiana 47907, USA}
\affiliation{Birck Nanotechnology Center, Purdue University, West Lafayette, Indiana 47907, USA}
\author{G.\,C. Gardner}
\affiliation{Birck Nanotechnology Center, Purdue University, West Lafayette, Indiana 47907, USA}
\affiliation{School of Materials Engineering, Purdue University, West Lafayette, Indiana 47907, USA}
\author{M.\,J. Manfra}
\affiliation{Department of Physics and Astronomy, Purdue University, West Lafayette, Indiana 47907, USA}
\affiliation{Birck Nanotechnology Center, Purdue University, West Lafayette, Indiana 47907, USA}
\affiliation{School of Materials Engineering, Purdue University, West Lafayette, Indiana 47907, USA}
\affiliation{School of Electrical and Computer Engineering, Purdue University, West Lafayette, Indiana 47907, USA}

\begin{abstract}
We investigate the effect of the filling factor on transport anisotropies, known as stripes, in high Landau levels of a two-dimensional electron gas.
We find that at certain in-plane magnetic fields, the stripes orientation is sensitive to the filling factor within a given Landau level.
This sensitivity gives rise to the emergence of stripes \emph{away from half-filling} while an orthogonally-oriented, native stripes reside at half-filling.
This switching of the anisotropy axes within a single Landau level can be attributed to a strong dependence of the native symmetry breaking potential on the filling factor. 
\end{abstract}
\maketitle

\begin{figure*}[t]
\centering
\includegraphics{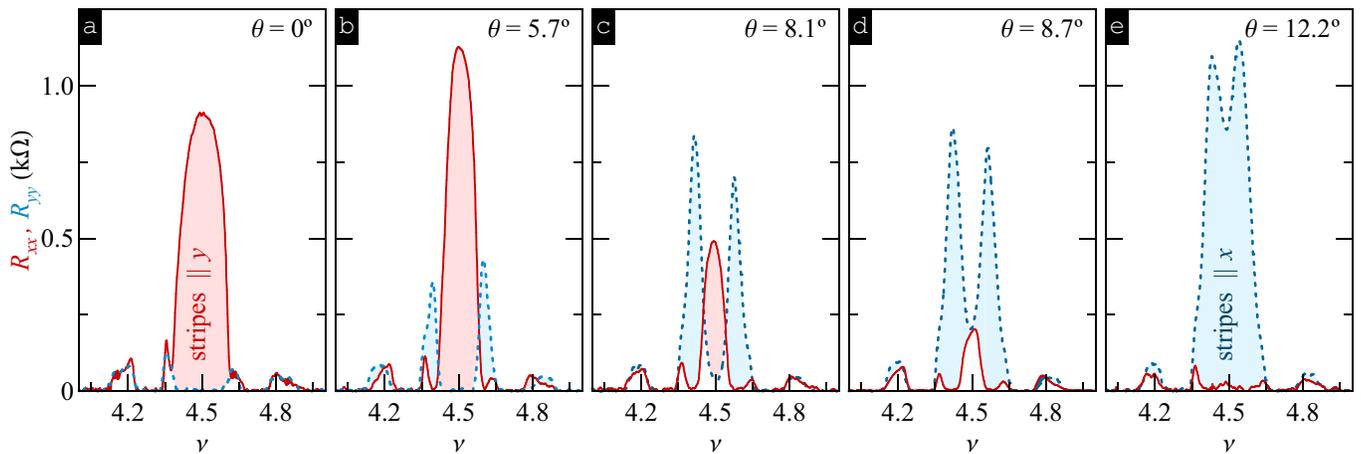}
\vspace{-0.15 in}
\caption{(Color online)
$\rxx$ (solid line) and $\ryy$ (dotted line) versus $\nu$ at $T = 25$ mK and $\theta$ from (a) $0^\circ$ to (e) $12.2^\circ$, as marked.
}
\vspace{-0.15 in}
\label{fig1}
\end{figure*}
Electronic analogs of liquid crystals, commonly termed electron nematics or stripes, are believed to exist in a wide variety of condensed matter systems \citep{fradkin:2010}, such as ruthenates \citep{borzi:2007}, high temperature superconductors \citep{daou:2010,chu:2010}, and heavy fermion systems \citep{okazaki:2011}. 
However, the first evidence for such nematic phases \cite{lilly:1999a,du:1999,koulakov:1996,fogler:1996,fradkin:1999,fradkin:2000} came from a clean 2D electron gas (2DEG) formed at the interface between GaAs and AlGaAs, which is often the system of choice to study of quantized Hall effects \citep{klitzing:1980,tsui:1982b}. 

One important parameter which is unique to a 2DEG, is the filling factor, $\nu = n_eh/eB$, where $n_e$ is the electron density and $B$ is the magnetic field.
Stripes in GaAs manifest themselves in the resistance minima (maxima) in the easy (hard) transport direction when $\nu$ is close to $\nu_{N}^\pm \equiv 2N + 1 \pm 1/2$ ($N \geq 2$), where $+$ ($-$) describes spin-down (spin-up) branches of the $N$-th Landau level.
While stripes exist in a finite filling factor range \citep{lilly:1999a,du:1999,sambandamurthy:2008,kukushkin:2011,note:5}, $\nu_N^\pm - 0.1 \lesssim \nu \lesssim \nu_N^\pm + 0.1$, transport studies have focused almost exclusively \citep{note:4} at half-integer $\nu$ where the anisotropy is the strongest.

In a purely perpendicular magnetic field, stripes orient along $\easy$ crystal direction with very few exceptions \citep{zhu:2002,pollanen:2015}. 
Despite nearly two decades of investigations, the origin of the native symmetry breaking potential responsible for this particular stripes orientation continues to remain elusive \cite{sodemann:2013,kovudayur:2011,pollanen:2015}.
It is well documented that an in-plane magnetic field $\bip$ provides an external symmetry breaking potential which competes with and can overcome the native symmetry breaking potential.
In particular, when applied along stripes (easy direction), $\bip$ can switch the anisotropy axes \cite{lilly:1999b,pan:1999,cooper:2001,jungwirth:1999,stanescu:2000,zhu:2009,pollanen:2015} aligning stripes perpendicular to it \citep{note:7}.
There thus exists a characteristic in-plane magnetic field $B_c$ which renders 2DEG macroscopically isotropic.
This field is routinely used to quantify the strength of the native symmetry breaking potential \citep{cooper:2001,cooper:2004,zhu:2002,pollanen:2015}.

In this Rapid Communication we report on transport studies of stripes in a high-mobility 2DEG focusing on filling factors \emph{away} from half-filling.
At certain $\bip$, we observe distinct anisotropic phases which reside both \emph{at} and \emph{away} from half-filling within a single Landau level.
These stripe phases have orthogonal orientation, indicating that it is largely determined by the filling factor.
The observed reorientations of stripes within a single Landau level implies strong sensitivity of the native symmetry breaking potential to the filling factor. 
Indeed, the characteristic in-plane field $B_c$ drops roughly symmetrically when $\nu$ deviates from half-filling in either direction.  
This dependence on the filling factor is quite significant and should be accounted for by theories attempting to identify the origin of the native symmetry breaking potential.

While similar results have been obtained from samples fabricated from several different wafers, here we present the data from one $\sim 4\times4$ mm square sample cleaved from a symmetrically doped, 30 nm-wide GaAs/AlGaAs quantum well. 
Electron density and mobility were $n_e \approx 2.9 \times 10^{11}$ cm$^{-2}$ and $\mu \approx 1.6 \times 10^7$ cm$^2$/Vs, respectively \citep{manfra:2014,note:21}.
Eight indium ohmic contacts were fabricated at the corners and midsides of the sample. 
The longitudinal resistances, $\rxx$ and $\ryy$, were measured using four-terminal, low-frequency lock-in technique; the current (typically 50 nA) was sent through the midside contacts and the voltage drop was measured between the corner contacts.
An in-plane magnetic field was introduced by tilting the sample by angle $\theta$ about $\x$ axis, i.e $\bip  = B_y = B \sin \theta$.
Unless otherwise noted, all the data we acquired at $T \approx 20$ mK.

In \rfig{fig1} we present $\rxx$ (solid line) and $\ryy$ (dotted line) versus filling factor $\nu$ at different $\theta$, from $0^\circ$ to $12.2^\circ$, as marked.
At $\theta  = 0^\circ$, the data reveal strong anisotropy near $\nu = 9/2$ with $\rxx \gg \ryy$ [see \rfig{fig1}(a)], i.e., stripes are oriented along $\y$ direction.
Near $\nu \approx 4.28$ and $\nu \approx 4.72$, $\rxx$ and $\ryy$ reveal isotropic insulating states, reflecting the formation of ``bubble'' phases \citep{lilly:1999a,du:1999,eisenstein:2002,gores:2007,deng:2012a,deng:2012b}.

At $\theta = 5.7^\circ$ [\rfig{fig1}(b)],
new anisotropic states of \emph{orthogonal} orientation (e.g., along $\x$-direction) develop near $\nu \approx 4.4$ and $\nu \approx 4.6$, i.e., at the edges of the native stripe range found at $\theta = 0^\circ$.
Like the native stripes, which reside near $\nu = 9/2$ \citep{note:3}, these new anisotropic states are represented by high peaks along one direction and deep minima along another, albeit with $\ryy \gg \rxx$.
As $\theta$ increases to 8.1$^{\circ}$ [\rfig{fig1}(c)], these orthogonal anisotropic states take over larger range of $\nu$, while the native anisotropy at $\nu = 9/2$ becomes noticeably weaker. 
At $\theta = 8.7^\circ$ [\rfig{fig1}(d)], stripes along $\x$-direction occupy almost the whole filling factor range of the native stripes, except near $\nu=9/2$, where the system becomes isotropic.
Finally, at $\theta = 12.2^{\circ}$ we observe that the $\bip$-induced reorientation of the native stripes is complete at all $\nu$ within the native stripe range.
The isotropic insulating states representing ``bubble'' phases remain essentially unchanged.

\begin{figure}[b]
\includegraphics{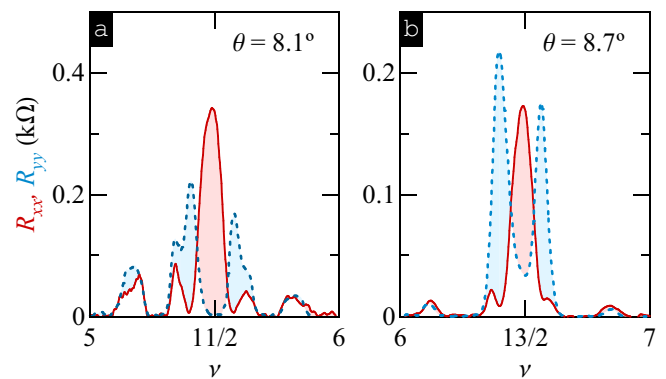}
\vspace{-0.15 in}
\caption{(Color online)
$\rxx$ (solid line) and $\ryy$ (dotted line) versus $\nu$ around (a) $\nu = 11/2$, $\theta = 8.1^\circ$ and (b) $\nu = 13/2$, $\theta = 8.7^\circ$.
}
\label{fig2}
\end{figure}

Before proceeding with the more detailed data analysis, we demonstrate that the evolution shown in \rfig{fig1} is not unique to $\nu = 9/2$.
In \rfig{fig2} we show $\rxx$ (solid line) and $\ryy$ (dotted line) at (a) $5 \le \nu \le 6$ and (b) $6 \le \nu \le 7$ measured at $\theta = 8.1^\circ$ and 8.7$^\circ$, respectively.
In both cases, the data show alternating stripes orientation which is identical to that shown in \rfig{fig1}(c). 
We thus conclude that the sensitivity of the stripes orientation to the filling factor is a generic feature of stripes in our sample; it appears in both spin-up and spin-down branches and in different ($N = 2, 3$) Landau levels.

\begin{figure*}[t]
\centering
\includegraphics{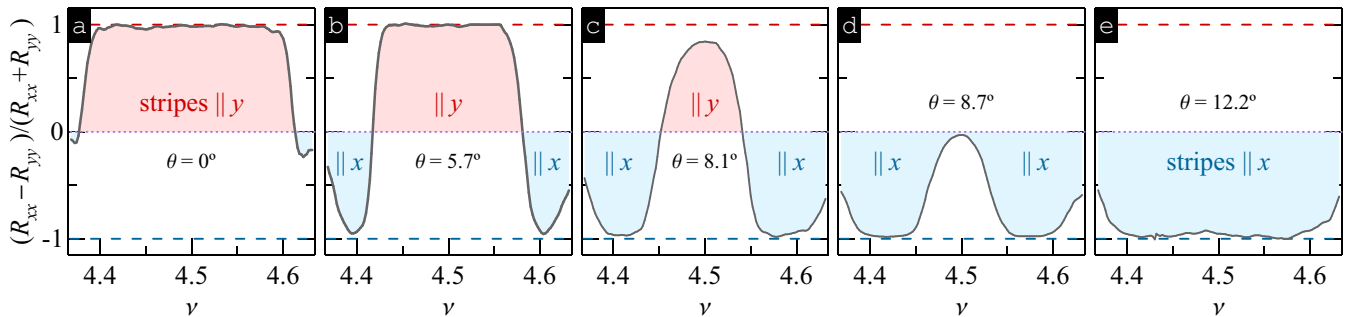}
\vspace{-0.15 in}
\caption{(Color online)
Resistance anisotropy $\A=(\rxx-\ryy)/(\rxx+\ryy)$ versus $\nu$ computed from the data in \rfig{fig1}.
}
\label{fig3}
\end{figure*}
To further examine the evolution of the anisotropic states with the tilt angle, we compute the resistance anisotropy $\A = (\rxx-\ryy)/(\rxx+\ryy)$ from the data shown in \rfig{fig1} and present the results in \rfig{fig3}.
At $\theta=0$ [\rfig{fig3}(a)] we see that $\A\approx 1$ at all filling factors within a band $4.4 \le \nu \le 4.6$, reflecting the native stripes orientation along $\y$ direction.
As the sample is tilted, this $\A \approx 1$ band  becomes narrower [\rfig{fig3}(b) and \rfig{fig3}(c)], and eventually vanishes as the system becomes isotropic at $\nu = 9/2$ where $\A \approx 0$ [\rfig{fig3}(d)].
Concomitant with the narrowing of the $\A \approx 1$ band (stripes $||\y$), we observe the emergence of the $\A \approx - 1$ bands (stripes $||\x$), which germinate at the edges of the native stripe band [\rfig{fig3}(b)].
With increasing $\theta$, these $\A \approx - 1$ bands expand towards each other [\rfig{fig3}(c)] and eventually merge at half-filling [\rfig{fig3}(d)]].
At the $\theta = 12.2^\circ$ [\rfig{fig3}(e)], the $\A\approx -1$ band occupies the whole range of the native stripes, $4.4 \le \nu \le 4.6$, reflecting stripes orientation along $\x$ direction.

Taken together, the data of \rfig{fig1}, \rfig{fig2}, and \rfig{fig3} demonstrate that the reorientation of stripes by $\bip$ depends sensitively on the filling factor \citep{note:6} and that at a certain fixed $\theta$ stripes orientation can vary within a single Landau level.
More specifically, we establish that the reorientation first occurs at filling factors near the onset of the native stripes, i.e. at $\nu \approx 4.4$ and $\nu \approx 4.6$, and then the boundary, separating two orthogonal stripe phases, propagates towards half-filling.

\begin{figure}[b]
\includegraphics{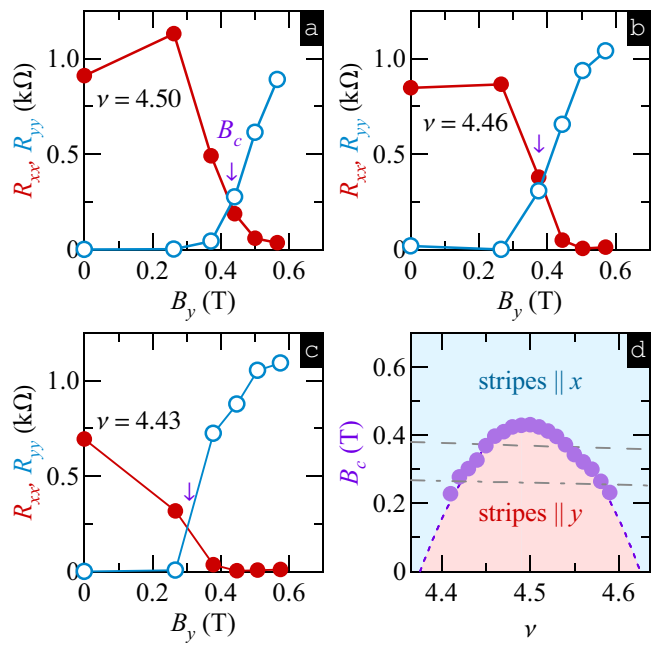}
\vspace{-0.15 in}
\caption{(Color online)
$\rxx$ (filled circles) and $\ryy$ (open circles) versus $B_y$ (a) $\nu = 4.5$, (b) 4.46 and (c) 4.43. 
(d) $B_c$ (filled circles) versus $\nu$. 
Dashed (dash-dotted) line represents $B_y$ at $\theta = 8.1^\circ$ ($\theta = 5.7^\circ$). 
}
\label{fig4}
\end{figure}

We next examine the evolution of the anisotropy with $\bip$ at different fixed filling factors in the vicinity of $\nu = 9/2$.
In \rfig{fig4}(a), (b), and (c), we present $\rxx$ (filled circles) and $\ryy$ (open circles) as a function of $B_y$, at $\nu$ = 4.5, 4.46, and 4.43, respectively.
In all three cases, we find that the anisotropy axes are switched when $\bip$ reaches a characteristic field $B_c$, defined such that $\rxx \approx \ryy$ (cf. $\downarrow$).
We further notice that $B_c$ \emph{decreases} as $\nu$ deviates from half-filling. 
We summarize these findings in \rfig{fig4}(d) showing $B_c$ as a function of $\nu$.
The dashed and dash-dotted lines in \rfig{fig4}(d) represent $\bip$ at $\theta= 8.1^\circ$ and $\theta = 5.7^\circ$, corresponding to the situation of \rfig{fig1}(c) and \rfig{fig1}(b), respectively.
Two crossings of these lines with $B_c$  is reflected in the transport data as switching of the anisotropy axes during the magnetic field sweep.

The characteristic field $B_c$ reaches its maximum value of about 0.43 T at $\nu = 9/2$.
As the filling factor deviates from half-filling, $B_c$ gradually decreases to $\approx 0.23$ T  at both $\nu = 4.41$ or $\nu = 4.59$ \citep{note:1}.
The dependence of $B_c$ on $\nu$ is roughly symmetric about half-filling and can be reasonably well described by a parabola (dotted line).
This parabola crosses zero at filling factors which are very close (within 0.01) to the onset of the native anisotropy [cf. \rfig{fig3}(a)].

Since $B_c$ characterizes the competition between the native symmetry breaking potential and the effect of $\bip$ on the orientation of stripes, the decrease of $B_c$ away from half-filling can be due to either a stronger effect of $\bip$ or a weaker native symmetry breaking potential at these filling factors.
However, since the effect of $\bip$ is predicted to have a very weak dependence on $\nu$ \cite{jungwirth:1999,stanescu:2000}, the relationship between $B_c$ and $\nu$ must reflect a weaker native symmetry breaking potential away from half-filling.
We note that the observed dependence is quite significant; $B_c$ decreases by nearly 50\% as $\nu$ is changed from 9/2 to 4.4  (or 4.6).

\begin{figure}[b]
\includegraphics{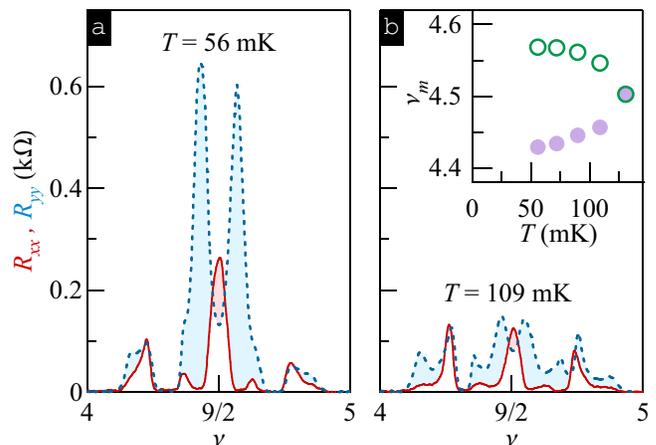}
\vspace{-0.15 in}
\caption{(Color online)
$\rxx$ (solid line) and $\ryy$ (dotted line) versus $\nu$ at $\theta = 8.5^{\circ}$ and (a) $T = 56$ mK and (b) 109 mK. 
The inset shows the change of $\nu_m$ at two $\ryy$ peaks versus $T$.
}
\label{fig5}
\end{figure}

Finally, we examine the temperature dependence of the anisotropic states away from half-filling.
In \rfig{fig5}(a) and \rfig{fig5}(b) we present $\rxx$ (solid line) and $\ryy$ (dotted line) measured at $\theta = 8.5^{\circ}$ and $T = 56$ mK and 109 mK, respectively.
As seen in \rfig{fig5}(a), $\ryy$ ($\rxx$) exhibits strong maxima (deep minima) away from half-filling and the corresponding anisotropy $\A \approx -1$ (stripes $||~\x$).
At $\nu = 9/2$, however, $\ryy$ exhibits a minimum, $\rxx$ a maximum, and the anisotropy becomes positive.
As the temperature is raised, the $\ryy$ peaks away from half-filling quickly decay, while the resistances at the corresponding $\rxx$ minima become larger [see \rfig{fig5}(b)].
As a result, the anisotropy away from half-filling is significantly reduced.
Interestingly, the $\ryy$ peaks also move towards each other with increasing $T$ and eventually merge into a single peak at half-filling before disappearing at higher temperatures.
This is illustrated in the inset of \rfig{fig5}(b) which shows the filling factors of the $\ryy$ peaks as a function of temperature. 
The merging of the $\ryy$ peaks indicates that the anisotropy away from half-filling disappears at a lower temperature compared to that at half-filling.

An interesting question is why the filling factor-dependence of $B_c$ has not been seen in previous experiments.
We argue that it is the significant native anisotropy away from half-filling, see \rfig{fig3}(a), that made possible a clear observation of this behavior.
Indeed, in the absence of an in-plane magnetic field our data show an unusually wide $\rxx$ peak and a vanishingly small $\ryy$ over a wide filling factor range, $4.4 \le \nu \le 4.6$, see \rfig{fig1}(a).
It is indeed this wide range which allows significant anisotropy (of the opposite sign) to develop away from half-filling at intermediate tilt angles [see \rfig{fig1}(b-d) and \rfig{fig3}(b-d)].
In contrast, in early studies \citep{lilly:1999b,pan:1999} the native anisotropy rapidly decays away from half-filling, as manifested by narrower resistance maxima and minima, and only weak signatures suggestive of filling factor-driven reorientation were observed [see, e.g. Fig. 2 in \rref{lilly:1999b}].

There is both experimental and theoretical evidence that weakening of the anisotropy away from half-filling is a result of stronger effect of disorder. 
Indeed, in agreement with the theory \cite{yi:2000}, pinning mode resonances in the ac conductivity \cite{sambandamurthy:2008} revealed higher resonance frequencies away from half-filling.
Although the sample in the present study has the mobility comparable to (or even lower than) samples used in earlier experiments, it is well established that the mobility alone is not a good metric of the quality of the transport data \citep{umansky:2009,gamez:2013,manfra:2014,samani:2014,watson:2015}.
We also notice that in contrast to early studies \citep{lilly:1999b,pan:1999}, which utilized conventional single heterointerfaces, our experiments exploit a double heterointerface of a ``doping well'' design \cite{manfra:2014}, which is known to produce better quality transport data revealed in, e.g., larger gaps of the fragile $\nu = 5/2$ fractional quantum Hall state \cite{deng:2014,samani:2014,watson:2015,note:8}. 

In addition to weak anisotropy away from half-filling there exist another argument why the phenomenon reported here can be suppressed in more disordered samples. 
Experiments on pinning mode resonances under $\bip$ \cite{zhu:2009} suggest that the disorder favors stripes \emph{parallel} to $\bip$.
This finding implies that due to stronger effect of disorder away from half-filling \cite{sambandamurthy:2008,yi:2000}, stripes at such filling factors would be more difficult to reorient perpendicular to $\bip$.
Thus, the effect of disorder would lead to \emph{larger $B_c$} away from half-filling which may mask the $\nu$-dependence of the native symmetry breaking field observed in the present study.

In summary, we have studied anisotropic phases in high Landau levels in tilted magnetic fields.
We have demonstrated the realization of anisotropic phases, distinct by their orthogonal orientation, which reside at and away from half-filling in a single Landau level \citep{note:101}.
The boundaries separating these states propagate towards each other and eventually merge at half-filling with increasing $\bip$.
The observed switching of the anisotropy axes within a single Landau level can be explained by a strong dependence of the native symmetry breaking potential on the filling factor. 
This conclusion is corroborated by a monotonic decrease of the characteristic in-plane magnetic field $B_c$ as $\nu$ deviates from half-filling.  
The experimentally observed dependence on the filling factor is quite significant and should be taken into account by theories attempting to identify the nature of the native symmetry breaking potential.

\begin{acknowledgments}
We thank G. Jones, S. Hannas, T. Murphy, J. Park, and A. Suslov for technical support.
We thank G. Csathy, L. Engel, H. Fertig, I. Kukushkin, M. Shayegan, and B. Shklovskii for discussions. 
The work at Minnesota (Purdue) was supported by the U.S. Department of Energy, Office of Science, Basic Energy Sciences, under Award \# ER 46640-SC0002567 (DE-SC0006671).
Q.S. acknowledges University of Minnesota Doctoral Dissertation Fellowship.
Experiments were performed at the National High Magnetic Field Laboratory, which is supported by NSF Cooperative Agreement No. DMR-0654118, by the State of Florida, and by the DOE.
\end{acknowledgments}


\begin{thebibliography}{57}
\bibitem[{not({\natexlab{e}})}]{note:21}
\bibinfo{note}{The doping wells are 90 nm away from the center of GaAs quantum
  well which is located 215 nm from the surface. At $T \approx 0.3$ K, the
  quantum lifetime obtained from the Dingle analysis of microwave-induced
  resistance oscillations \citep{zudov:2001a} is about 20 ps.}
\bibitem[{not({\natexlab{h}})}]{note:1}
\bibinfo{note}{We do not include $B_c$ at smaller or larger $\nu$ because of
  large experimental uncertainty in this regime.}
\end{thebibliography}

\end{document}